# An Effective Clustering Approach to Web Query Log Anonymization

Amin Milani Fard, and Ke Wang, *Senior Member, IEEE*

*Abstract*—Web query log data contain information useful to research; however, release of such data can re-identify the search engine users issuing the queries. These privacy concerns go far beyond removing explicitly identifying information such as name and address, since non-identifying personal data can be combined with publicly available information to pinpoint to an individual. In this work we model web query logs as unstructured transaction data and present a novel transaction anonymization technique based on clustering and generalization techniques to achieve the *k*-anonymity privacy. We conduct extensive experiments on the AOL query log data. Our results show that this method results in a higher data utility compared to the state-of-the-art transaction anonymization methods.

*Index Terms*— Query logs data, privacy-preserving data publishing, transaction data anonymization, item generalization.

## I. INTRODUCTION

WEB search engines generally store query logs data for the purpose of improving ranking algorithms, query refinement, user modeling, fraud/abuse detection, language-based applications, and sharing data for academic research or commercial needs [4]. On the other hand, the release of query logs data can seriously breach the privacy of search engine users. The privacy concern goes far beyond just removing the identifying information from a query. Sweeney [17] showed that even non-identifying personal data can be combined with publicly available information, such as census or voter registration databases, to pinpoint to an individual. In 2006 the America Online (AOL) query logs data, over a period of three months, was released to the public [2]. Although all explicit identifiers of searchers have been removed, by examining query terms, the searcher No. 4417749 was traced back to the 62-year-old widow Thelma Arnold. Since this scandal, data publishers become reluctant to provide researchers with public anonymized query logs [7].

An important research problem is how to render web query log data in such a way that it is difficult to link a query to a specific individual while the data is still useful to data analysis. Several recent works start to examine this problem, with [10] and [1] from web community focusing on privacy attacks, and [8], [18], and [19] from the database community focusing on anonymization techniques. Although good progresses are

This work was supported by a Natural Sciences and Engineering Research Council of Canada (NSERC) Discovery Grant.

A. Milani Fard and K. Wang are with the School of Computing Science, Simon Fraser University, 8888 University Drive, Burnaby, BC V5A 1S6, Canada. E-mail: {milanifard, wangk}@cs.sfu.ca.

made, a major challenge is reducing the significant information loss of the anonymized data.

The subject of this paper falls into the field of *privacy preserving data publishing* (*PPDP*) [6], which is different from access control and authentication associated with computer security. The work in these latter areas ensures that the recipient of information has the authority to receive that information. While such protections can safeguard against direct disclosures, they do not address disclosures based on inferences that can be drawn from released data. The subject of *PPDP* is not much on whether the recipient can access to the information or not, but is more on what values will constitute the information the recipient will receive so that the privacy of record owners is protected.

### A. Motivations

This paper studies the query log anonymization problem with the focus on reducing information loss. One approach is modeling query logs data as a special case of transaction data, where each transaction contains several "items" from an item universe *I*. In the case of query logs, each transaction represents a query and each item represents a query term. Other examples of transaction data are emails, online clicking streams, online shopping transactions, and so on. As pointed out in [18] and [19], for transaction data, the item universe I is very large (say thousands of items) and a transaction contains only a few items. For example, each query contains a tiny fraction of all query terms that may occur in a query log. If each item is treated as a binary attribute with 1/0 values, the transaction data is extremely high dimensional and sparse. On such data, traditional techniques suffer from extreme information loss [18] and [19].

Recently, the authors of [8] adapted the top-down *Mondrian* [12] partition algorithm originally proposed for relational data to generalize the set-valued transaction data. We refer to this algorithm as *Partition* in this paper. They adapted the traditional *k*-anonymity [15] and [16] to the set valued transaction data. A transaction database is *k-anonymous* if transactions are partitioned into equivalence classes of size at least *k*, where all transactions in the same equivalence class are exactly identical. This notion prevents linking attacks in the sense that the probability of linking an individual to a specific transaction is no more than 1/*k*.

Our insight is that *Partition* method suffers from significant information loss on transaction data. Consider the transaction data $S$={ $t_1$, $t_2$, $t_3$, $t_4$, $t_5$ } in the second column of Table I and the item taxonomy in Fig. 1. Assume $k = 2$. *Partition* works as follows. Initially, there is one partition $P_{\{food\}}$ in which the items in every transaction are generalized to the top-most item *food*. At this point, the possible drill-down is *food* $\rightarrow$ {*fruit*, *meat*, *dairy*}, yielding $2^3$-1 sub-partitions corresponding to the non-empty subsets of {*fruit*, *meat*, *dairy*}, i.e., $P_{\{fruit\}}$, $P_{\{meat\}}$, ..., and $P_{\{fruit,meat,dairy\}}$, where the curly bracket of each sub-partition contains the common items for *all* the transactions in that sub-partition. All transactions in $P_{\{food\}}$ are then partitioned into these sub-partitions. All sub-partitions except $P_{\{fruit,meat\}}$ violate *k*-anonymity (for *k*=2) and thus are merged into one partition $P_{\{food\}}$. Further partitioning of $P_{\{fruit,meat\}}$ also violates *k*-anonymity. Therefore, the algorithm stops with the result shown in the last column of Table I.

One drawback of *Partition* is that it stops partitioning the data at a high level of the item taxonomy. Indeed, specializing an item with $n$ children will generate $2^n$-1 possible sub-partitions. This exponential branching, even for a small value of $n$, quickly diminishes the size of a sub-partition and causes violation of *k*-anonymity. This is especially true for query logs data where query terms are drawn from a large universe and are from a diverse section of the taxonomy.

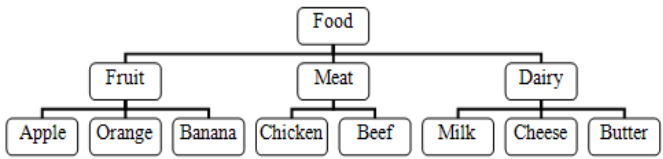

Fig. 1. Food taxonomy tree

TABLE I
The motivating example and its 2-anonymization

| TID | Original Data | Partition |
|-----|---------------|-----------|
| $t_1$ | *<orange, chicken, beef>* | *<fruit, meat>* |
| $t_2$ | *<banana, beef, cheese>* | *<food>* |
| $t_3$ | *<chicken, milk, butter>* | *<food>* |
| $t_4$ | *<apple, chicken>* | *<fruit, meat>* |
| $t_5$ | *<chicken, beef>* | *<food>* |

Moreover, the *Partition* does not deal with item duplication. As an example, the generalized $t_3$ in the third column of Table I contains only one occurrence of *food*, which clearly has more information loss than the generalized transaction *<food, food, food>* because the latter tells more truthfully that the original transaction purchases at least three items. Indeed, the TFIDF used by many ranking algorithms critically depends on the term frequency of a term in a query or document. Preserving the occurrences of items (as much as possible) would enable a wide range of data analysis and applications.

### B. Contributions

To render the input transaction data k-anonymous, our observation is: if "similar" transactions are grouped together, less generalization and suppression will be needed to render them identical. As an example, grouping two transactions *<Apple>* and *<Milk>* (each having only one item) entails more information loss than grouping two transactions *<Apple>* and *<Orange>*, because the former results in the more generalized transaction *<Food>* whereas the latter results in the less generalized transaction *<Fruit>*. Therefore, with a proper notion of transaction similarity, we can treat the transaction anonymization as a clustering problem such that each cluster must contain at least k transactions and these transactions should be "similar". Our main contributions are as follows:

**Contribution 1** For a given item taxonomy, we introduce the notion of the *Least Common Generalization* (*LCG*) as the generalized representation of a subset of transactions, and as a way to measure the similarity of a subset of transactions. The distortion of *LCG* models the information loss caused by both item generalization and item suppression. We devise a linear-time algorithm to compute *LCG*.

**Contribution 2** We formulate the transaction anonymization as the problem of clustering a given set of transactions into clusters of size at least *k* such that the sum of *LCG* distortion of all clusters is minimized.

**Contribution 3** We present a heuristic linear-time solution to the transaction anonymization problem.

**Contribution 4** We evaluate our method on the AOL query logs data.

The structure of the paper is as follows. Section II describes problem statements. Section III gives our clustering algorithm. Section IV presents the detailed algorithm for computing *LCG*. Section V presents the experimental results. Section VI reviews related works. We conclude in Section VII.

## II. PROBLEM STATEMENTS

This section defines our problems. We use the terms "transaction" and "item". In the context of web query logs, a transaction corresponds to a query and an item corresponds to a query term.

### A. Item Generalization

We assume that there is a taxonomy tree $T$ over the item universe $I$, with the parent being more general than all children. This assumption was made in the literature [15], [16], [8], [18]. For example, *WordNet* [5] could be a source to obtain the item taxonomy.

The process of generalization refers to replacing a special item with a more general item (i.e., an ancestor), and the process of specialization refers to the exact reverse operation. In this work, an item is its own ancestor and descendant.

**Definition 1** (***Transactions and generalization***) A *transaction* is a bag of items from $I$ (thus allowing duplicate items). A transaction $t'$ is a *Generalized Transaction* of a transaction $t$, if for every item $i' \in t'$ there exists one *distinct* item $i \in t$ such that $i'$ is an ancestor of $i$. In this case, $t$ is the *Specialized Transaction* of $t'$. □

The above transaction model is different from [8] in several ways. First, it allows duplicate items in a transaction. Second, it allows items in a transaction to be on the same path in the item taxonomy, in which case, each item represents a distinct leaf item. For example, we interpret the transaction $<Fruit, Food>$ as: *Fruit* represents (the generalization of) a leaf item under *Fruit* and *Food* represents a leaf item under *Food* that is not represented by *Fruit*. Also, if $t'$ is a generalized transaction of $t$, each item $i' \in t'$ represents one *distinct* item $i \in t$. We say that an item $i \in t$ is *suppressed* in $t'$ if no $i' \in t'$ represents the item $i$. Hence, our generalization also models item suppression.

**Example 1** Consider the taxonomy tree in Fig. 1 and the transaction $t=<Orange, Beef>$. All possible generalized transactions of $t$ are $<>$, $<Orange>$, $<Beef>$, $<Orange, Beef>$, $<Fruit, Beef>$, $<Orange, Meat>$, $<Fruit, Meat>$, $<Fruit>$, $<Meat>$, $<Food>$, $<Orange, Food>$, $<Food, Beef>$, $<Fruit, Food>$, $<Food, Meat>$, and $<Food, Food>$. For $t'=<Fruit>$, *Fruit* represents (the generalization of) some item under the category *Fruit* (i.e., *Orange*), and *Beef* is a suppressed item since no more item in $t'$ represents it. For $t'=<Food>$, *Food* represents one item under *Food*, therefore, one of *Orange* and *Beef* in $t$ is suppressed. For $t'=<Food, Food>$, each occurrence of *Food* represents a different item in $t$. □

### B. Least Common Generalization

The main idea of transaction anonymization is to build groups of identical transactions through generalization. We introduce the following notion to capture such generalizations.

**Definition 2 (LCG)** The *Least Common Generalization* of a set of transactions $S$, denoted by $LCG(S)$, is a common generalized transaction for all transactions in $S$, and there is no other more special common generalized transaction. □

The following properties follow from the above definition. The proof has been omitted due to the space limit.

**Property 1** $LCG(S)$ is unique for a given $S$.

**Property 2** The length of $LCG(S)$ (i.e. the number of items in it) is equal to the length of the shortest transaction in $S$. This property can be ensured by padding the root item to $LCG$ if necessary.

**Example 2** Consider the taxonomy tree in Fig. 1. Let $S_1 = \{<Orange, Beef>, <Apple, Chicken, Beef>\}$, $LCG(S_1) = <Fruit, Beef>$. $LCG(S_1)$ cannot be $<Fruit, Meat>$ since $<Fruit, Beef>$ is a more specialized common transaction. For $S_2 = \{<Orange, Milk>, <Apple, Cheese, Butter>\}$, $LCG(S_2)=<Fruit, Dairy>$. *Dairy* represents *Milk* in the first transaction and represents one of *Cheese* and *Butter* in the second transaction. Thus one of *Cheese* or *Butter* is considered as a suppressed item. For $S_3 = \{<Orange, Apple>, <Orange, Banana, Milk>, <Banana, Apple, Beef>\}$, $LCG(S_3)=<Fruit, Fruit>$, which represents that all three transactions contain at least two items under *Fruit*. *Milk* and *Beef* are suppressed items. For $S_4 = \{<Orange, Beef>, <Apple, Milk>\}$, $LCG(S_4) = <Fruit, Food>$, where *Food* represents *Beef* in the first transaction and *Milk* in the second transaction. Here *LCG* contains both a parent and a child item. □

Various metrics have been proposed in the literature to measure the quality of generalized data including *Classification Metric* (*CM*), *Generalized Loss Metric* (*LM*) [9], and *Discernibility Metric* (*DM*) [3]. We use *LM* to measure item generalization distortion. The similar notion of *NCP* has also been employed for set-valued data [18] and [8]. Let $M$ be the total number of leaf nodes in the taxonomy tree $T$, and let $Mp$ be the number of leaf nodes in the subtree rooted at a node $p$. The *Loss Metric* for an item $p$, denoted by $LM(p)$, is defined as $(Mp-1) / (M-1)$. For the root item $p$, $LM(p)$ is 1. In words, *LM* captures the degree of generalization of an item by the percentage of the leaf items in the domain that are indistinguishable from it after the generalization. For example, considering taxonomy in Fig. 1, $LM(Fruit)=2/7$.

Suppose that we generalize every transaction in a subset of transactions $S$ to a common generalized transaction $t$, and we want to measure the distortion of this generalization. Recall that every item in $t$ represents one *distinct* item in each transaction in $S$ (Definition 1). Therefore, each item in $t$ generalizes exactly $|S|$ items, one from each transaction in $S$, where $|S|$ is the number of transactions in $S$. The remaining items in a transaction (that are not generalized by any item in $t$) are suppressed items. Therefore, the distortion of this generalization is the sum of the distortion for generalized items, $|S| \times \Sigma_{i \in t} LM(i)$, and the distortion for suppressed items. For each suppressed item, we charge the same distortion as if it is generalized to the *root* item, i.e., 1.

**Definition 3 (GGD)** Suppose that we generalize every transaction in a set of transactions $S$ to a common generalized transaction $t$. The *Group Generalization Distortion* of the generalization is defined as $GGD(S, t) = |S| \times \Sigma_{i \in t} LM(i) + N_s$, where $N_s$ is the number of occurrences of suppressed items. □

To minimize the distortion, we shall generalize $S$ to the least common generalization $LCG(S)$, which has the distortion $GGD(S, LCG(S))$.

**Example 3** Consider the taxonomy in Fig. 1 and $S_1=\{<Orange, Beef>, <Apple, Chicken, Beef>\}$. We have $LCG(S_1) = <Fruit, Beef>$. $LM(Fruit)=2/7$, $LM(Beef)=0$, and $|S_1|=2$. Since *Chicken* is the only suppressed item, $N_s=1$. Thus $GGD(S_1, LCG(S_1)) = 2 \times (2/7+0) + 1 = 11/7$. □

### C. Problem Definition

We adopt the transactional $k$-anonymity in [8] as our privacy notion. A transaction database $D$ is $k$-*anonymous* if for every transaction in $D$, there are at least $k-1$ other identical transactions in $D$. Therefore, for a $k$-anonymous $D$, if one transaction is linked to an individual, so are at least $k-1$ other transactions, so the adversary has at most $1/k$ probability to link a specific transaction to the individual. For example, the last column in Table I is a 2-anonymous transaction database.

**Definition 4 (Transaction anonymization)** Given a transaction database $D$, a taxonomy of items, and a privacy parameter $k$, we want to find the clustering $C=\{S_1,...,S_n\}$ of $D$ such that $S_1,...,S_n$ are pair-wise disjoint subsets of $D$ with each $S_i$ containing at least $k$ transactions from $D$, and $\Sigma_{i=1..|C|} GGD(S_i, LCG(S_i))$ is minimized. □

Let $C=\{S_1,...,S_n\}$ be a solution to the above anonymization problem. A $k$-anonymized database of $D$ can be obtained by generalizing every transaction in $S_i$ to $LCG(S_i)$, $i=1,...,n$.

## III. CLUSTERING APPROACH

In this section we present our algorithm *Clump* for solving the problem defined in Definition 4. In general, the problem of finding optimal *k*-anonymization is *NP*-hard for $k \geq 3$ [13]. Thus, we focus on an efficient heuristic solution to this problem and evaluate its effectiveness empirically. In this section, we assume that the functions $LCG(S)$ and $GGD(S, LCG(S))$ are given. We will discuss the detail of computing these functions in Section IV.

The central idea of our algorithm is to group transactions in order to reduce $\Sigma GGD(S_i, LCG(S_i))$, subject to the constraint that $S_i$ contains at least *k* transactions. Recall $GGD(S, LCG(S)) = |S| \times \Sigma_{i \in LCG(S)} LM(i) + N_s$ and from Property 2, $LCG(S)$ has the length equal to the minimum length of transactions in *S*. All "extra" items in a transaction that do not have a generalization in $LCG(S)$ are suppressed and contributes to the suppression distortion $N_s$. Since the distortion of generalizing an item is no more than the distortion of suppressing an item, one heuristic is to group transactions of similar length into one cluster in order to minimize the suppression distortion $N_s$.

Based on this idea, we present our algorithm *Clump*. Let *D* be the input transaction database and let $n = \lfloor |D|/k \rfloor$ be the number of clusters, where $|D|$ denotes the number of transactions in *D*:

**Step 1** (line 2-5): We arrange the transactions in *D* in the decreasing order of the transaction length, and we initialize the $i^{th}$ cluster $S_i$, $i=1,...,n$, with the transaction at the position $(i-1)k+1$ in the ordered list. Since earlier transactions in the arranged order have longer length, earlier clusters in this order tend to contain longer transactions.

For the comparison purpose, we also implement other transaction assignment orders, such as random assignment order and the increasing transaction length order (i.e., the exact reverse order of the above algorithm). Our experiments found that the decreasing order by transaction length produced better results.

**Step 2** (line 6-12): For each remaining transaction $t_i$ in the arranged order, we assign $t_i$ to the cluster $S_j$ such that $|S_j| < k$ and $GGD(S_j \cup \{t_i\}, LCG(S_j \cup \{t_i\}))$ is minimized. Since this step requires computing $GGD(S_j \cup \{t_i\}, LCG(S_j \cup \{t_i\}))$, we can restrict the search to the first *r* clusters $S_j$ with $|S_j| < k$, where *r* is a pruning parameter. Our order of examining transactions implies that longer transactions tend to be assigned to earlier clusters.

**Step 3** (line 13-17): after all of the *n* clusters contain *k* number of transactions, for each remaining transaction $t_i$ in the sorted order, we assign it to the cluster $S_j$ with the minimum $GGD(S_j \cup \{t_i\}, LCG(S_j \cup \{t_i\}))$.

The major work of the algorithm is computing $GGD(S_j \cup \{t_i\}, LCG(S_j \cup \{t_i\}))$, which requires the $LCG(S_j \cup \{t_i\})$. We will present an algorithm for computing $LCG(S_i)$ in time $O(|T| \times |S_i|)$ in the next section, where $|T|$ is the size of the taxonomy tree *T* and $|S_i|$ is the number of transactions in $S_i$. It is important to note that each cluster $S_i$ has a size at most 2*k*. Since *k* is small, *LCG* can be computed efficiently. In fact, the next lemma says that $LCG(S_j \cup \{t_i\})$ can be computed incrementally from $LCG(S_j)$.

---

**Algorithm 1** *Clump*: Transaction Clustering

**Input:** Transaction database: *D*, Taxonomy: *T*, Anonymity parameter: *k*, $n = \lfloor |D|/k \rfloor$
**Output:** *k*-anonymous transaction database: *D\**
**Method:**
1. Initialize $S_i \leftarrow \varnothing$ for i=1,...,|D|;
2. Sort the transactions in *D* in the descending order of length
3. **for** $i = 1$ to n **do**
4.     assign the transaction at the position $(i-1)k+1$ to $S_i$
5. **end for**
6. **while** $|S_j| < k$ for some $S_j$ **do**
7.     **for** each unassigned transaction $t_i$ in sorted order **do**
8.         Let $S_j$ be the cluster such that $|S_j| < k$ and $GGD(S_j \cup \{t_i\}, LCG(S_j \cup \{t_i\}))$ is minimized
9.         $LCG(S_j) \leftarrow LCG(S_j \cup \{t_i\})$
10.         $S_j \leftarrow S_j \cup \{t_i\}$
11.     **end for**
12. **end while**
13. **for** each unassigned transaction $t_i$ **do**
14.     Let $S_j$ be the cluster such that $GGD(S_j \cup \{t_i\}, LCG(S_j \cup \{t_i\}))$ is minimized
15.     $LCG(S_j) \leftarrow LCG(S_j \cup \{t_i\})$
16.     $S_j \leftarrow S_j \cup \{t_i\}$
17. **end for**
18. return $LCG(S_i)$ and $S_i$, i=1,..., n

---

**Lemma 1** Let *t* be a transaction, *S* be a subset of transactions, and $S'=\{LCG(S), t\}$ consist of two transactions. Then $LCG(S \cup \{t\}) = LCG(S')$.

*Proof: Omitted due to the space limit.* □

In words, the lemma says that the *LCG* of $S_j \cup \{t_i\}$ is equal to the *LCG* of two transactions, $LCG(S)$ and $t_i$. Thus if we maintain $LCG(S_j)$ for each cluster $S_j$, the computation of $LCG(S_j \cup \{t\})$ involves only two transactions and takes the time $O(|T|)$.

**Theorem 1** For a database *D* and a taxonomy tree *T*, Algorithm 1 runs in time $O(|D| \times r \times |T|)$, where *r* is the pruning parameter used by the algorithm.

*Proof:* We apply *Counting Sort* which takes $O(|D|)$ time to sort all transactions in *D* by their length. Subsequently, the algorithm examines each transaction once to insert it to a cluster. To insert a transaction $t_i$, the algorithm examines *r* clusters and, for each cluster $S_j$, it computes $LCG(S_j \cup \{t_i\})$ and $GGD(S_j \cup \{t_i\}, LCG(S_j \cup \{t_i\}))$, which takes $O(|T| \times |S_j|)$ according to Theorem 2 in Section IV, where $|S_j|$ is the number of transactions in $S_j$. With the incremental computing of $LCG(S_j \cup \{t_i\})$ in Lemma 1, computing $LCG(S_j \cup \{t_i\})$ takes time proportional to $|T|$. Overall, the algorithm is in $O(|D| \times r \times |T|)$. □

Since $|T|$ and *r* are constants, the algorithm takes a linear time in the database size $|D|$.

## IV. COMPUTING LCG

In the previous section, we make use of the functions $LCG(S)$ and $GGD(S, LCG(S))$ to determine the cluster for a transaction. Since these functions are frequently called, an efficient implementation is crucial. In this section, we present a linear time algorithm for computing $LCG$ and $GGD$. We focus on $LCG$ because computing $GGD$ is straightforward once $LCG$ is found.

### A. Bottom-Up Generalization

We present a bottom-up item generalization ($BUIG$) algorithm to build $LCG(S)$ for a set $S$ of transactions. First, we initialize $LCG(S)$ with the empty set of items. Then, we examine the items in the taxonomy tree $T$ in the bottom-up fashion: examine a parent only after examining all its children. For the current item $i$ examined, if $i$ is an ancestor of some item in *every* transaction in $S$, we add $i$ to $LCG(S)$. In this case, $i$ is the least common generalization of these items. If $i$ is not an ancestor of any item in some transaction in $S$, we need to examine the parent of $i$.

This algorithm is described in Algorithm 2. Let $S = \langle t_1,\ldots,t_m\rangle$. For an item $i$, we use an array $R_i[1..m]$ to store the number of items in a transaction of which $i$ is an ancestor. Specifically, $R_i[j]$ is set to the number of items in the transaction $t_j$ of which $i$ is an ancestor. $MinCount(R_i)$ returns the minimum entry in $R_i$, i.e., $\min_{j=1..m} R_i[j]$. If $MinCount(R_i)>0$, $i$ is an ancestor of at least $MinCount(R_i)$ distinct items in *every* transaction in $S$, so we will add $MinCount(R_i)$ copies of the item $i$ to $LCG(S)$.

Algorithm 2 is a call to the recursive procedure $BUIG(root)$ with the *root* of $T$. Line 1-6 in the main procedure initializes $LCG$ and $R_i$. Consider $BUIG(i)$ for an item $i$. If $i$ is a leaf in $T$, it returns. Otherwise, line 4-9 examines recursively the children $i'$ of $i$, by the call $BUIG(i')$. On return from $BUIG(i')$, if $MinCount(R_{i'})>0$, $i'$ is an ancestor of at least $MinCount(R_{i'})$ items in *every* transaction in $S$, so $MinCount(R_{i'})$ copies of $i'$ are added to $LCG$. If $MinCount(R_{i'})=0$, $i'$ does not represent any item for some transaction in $S$, so the examination moves up to the parent item $i$; in this case, line 8 computes $R_i$ by aggregating $R_{i'}$ for all child items $i'$ such that $MinCount(R_{i'})=0$. Note that, by not aggregating $R_{i'}$ with $MinCount(R_{i'})>0$, we stop generalizing such child items. If $i$ is the *root*, line 10-11 adds $MinTranSize(S)-|LCG|$ copies of the *root* item to $LCG$, where $MinTranSize(S)$ returns the minimum transaction length of $S$. This step ensures that $LCG$ has the same length as the minimum transaction length of $S$ (Property 2).

**Example 5** Let $S=\{\langle Orange, Apple\rangle, \langle Orange, Banana, Milk\rangle, \langle Banana, Apple, Beef\rangle\}$ and consider the taxonomy in Fig. 1. $BUIG(Food)$ recurs until reaching leaf items. Then the processing proceeds bottom-up as depicted in Fig. 2. Next to each item $i$, we show $o:R_i$, where $o$ is the sequence order in which $i$ is examined and $R_i$ stores the number of items in each transaction of which $i$ is an ancestor.

The first three items examined are *Apple*, *Orange*, and *Banana*. $R_{Apple} = [1,0,1]$ (since *Apple* appeared in transactions 1 and 3), $R_{Orange}=[1,1,0]$, and $R_{Banana}=[0,1,1]$. $MinCount(R_i)=0$

for these items $i$. Next, the parent *Fruit* is examined and $R_{Fruit} = R_{Apple} + R_{Orange}+ R_{Banana}=[2,2,2]$. With $MinCount(R_{Fruit})$ = 2, two copies of *Fruit* are added to $LCG$, i.e., $LCG(S)=\langle Fruit, Fruit\rangle$ and we stop generalizing *Fruit*.

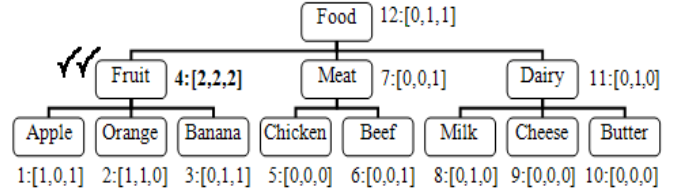

Fig. 2. BUIG's processing order

A similar processing applies to the sub-trees at *Meat* and *Dairy*, but no item $i$ is added to $LCG$ because $MinCount(R_i)=0$. Finally, at the root *Food*, $R_{Food} = R_{Meat} + R_{Dairy} = [0,1,1]$. Note that we do not add $R_{Fruit}$ because $MinCount(R_{Fruit})>2$, which signals that the generalization has stopped at *Fruit*. Since $|LCG|=MinTranSize(S)$, no *Food* is added to $LCG$. So the final $LCG(S)=\langle Fruit, Fruit\rangle$. As mentioned in Example 2, the two occurrences of *Fruit* indicate that all three transactions contain at least two items under *Fruit*. □

---

**Algorithm 2** Bottom-up Item Generalization

**Input:** Taxonomy: $T$, Set of $m$ transactions: $S = \langle t_1, ..., t_m\rangle$
**Output:** $LCG(S)$
**Method:**
1. $LCG \leftarrow \varnothing$;
2. **for each** item $i \in T$ **do**
3.      **for each** $t_j \in$ S **do**
4.          **if** $t_j$ contains $i$ **then** $R_i[j] \leftarrow 1$ **else** $R_i[j] \leftarrow 0$
5.      **end for**
6. **end for**
7. $BUIG(root)$;
8. **return** $LCG$;

                ****

$BUIG(i)$:
1. **if** $i$ is a leaf in $T$ **then**
2.      **return**
3. **else**
4.      **for each** child $i'$ of $i$ **do**
5.          $BUIG(i')$;
6.          **if** $MinCount(R_{i'})>0$ **then**
7.          Add $MinCount(R_{i'})$ copies of $i'$ to $LCG$
8.          **else** $R_i \leftarrow R_i+R_{i'}$    /* examining the parent $i$ */
9.      **end for**
10.      **if** $i=root$ **then**
11.      Add $MinTranSize(S)-|LCG|$ copies of $root$ to $LCG$
12. **return**

---

**Theorem 2** Given a set of transactions $S$ and a taxonomy tree $T$ of items, $BUIG$ produces $LCG(S)$ and takes time $O(|T| \times |S|)$, where $|S|$ is the number of transactions in $S$ and $|T|$ is the number of items in taxonomy tree $T$.

*Proof:* First, $BUIG$ generalizes transactions by examining the items in $T$ in the bottom-up order and stops generalization whenever encountering an item that is a common ancestor of some unrepresented item in every transaction in $S$. This property ensures that each item added to $LCD$ is the earliest possible common ancestor of some unrepresented item in every transaction. Second, *BUIG* visits each node in $T$ once, and at each node $i$, it examines the structures $R_{i'}$ and $R_j$ of size $|S|$, where $i'$ is a child of $i$. So the complexity is $O(|T| \times |S|)$. □

### B. A Complete Example

Let us illustrate the complete run of *Clump* using the motivating example in Section I.A. We reproduce the five transactions $t_1$ to $t_5$ in Table II, arranged by the descending order of transaction length. Let $k$=2. First, the number of clusters is $m = \lfloor 5/2 \rfloor = 2$, and the first cluster $S_1$ is initialized to the first transaction $t_1$ and the second cluster $S_2$ is initialized to the third transaction $t_3$. Next, we assign the remaining transactions $t_2$, $t_4$, and $t_5$ in that order. Consider $t_2$. If we assign $t_2$ to $S_1$, $LCG(S_1 \cup \{t_2\})=\{fruit,beef,food\}$, and $GGD = 2 \times (2/7+0+1) = 2.57$. If we assign $t_2$ to $S_2$, we have $LCG(S_2 \cup \{t_2\})=\{meat,dairy,food\}$ and $GGD = 2 \times (1/7+2/7+1) = 2.85$. Thus the decision is assigning $t_2$ to $S_1$ resulting in $S_1=\{t_1,t_2\}$ and $LCG(S_1)=\{fruit, beef, food\}$.

Next, we assign $t_4$ to $S_2$ because $S_1$ has contained $k$=2 transactions. So $S_2=\{t_3, t_4\}$ and $LCG(S_2)= \{chicken, food\}$. Next, we have the choice of assigning $t_5$ to $S_1$ or $S_2$ because both have contained 2 transactions. The decision is assigning $t_5$ to $S_2$ because it results in a smaller $GGD$, and $LCG(S_2)= <chicken,food>$. So the final clustering is $S_1=\{t_1, t_2\}$ and $S_2=\{t_3, t_4, t_5\}$. The last column of Table II shows the final generalized transactions.

#### Table II
#### The motivating example and its 2-anonymization

| ID | Original Data | Partition | Clump |
|---|---|---|---|
| $t_1$ | <orange,chicken,beef> | <fruit,meat> | <fruit,beef,food> |
| $t_2$ | <banana,beef,cheese> | <food> | <fruit,beef,food> |
| $t_3$ | <chicken,milk,butter> | <food> | <chicken,food> |
| $t_4$ | <apple,chicken> | <fruit,meat> | <chicken,food> |
| $t_5$ | <chicken,beef> | <food> | <chicken,food> |

Let us compare this result of *Clump* with the result of *Partition* in the third column (which has been derived in Section I.A). For *Clump*, we measure the distortion by $\Sigma GGD(S_i, LCG(S_i))$ over all clusters $S_i$. For *Partition*, we measure the distortion by $\Sigma GGD(S_i, t_j)$ over all sub-partitions $S_i$ where $t_j$ is the generalized transaction for $S_i$. The $GGD$ for *Clump* is $2 \times (2/7+0+1) + [3 \times (0+1)+1] = 6.57$, compared to $[2 \times (2/7+1/7)+1] + [3 \times (1)+5] = 8.85$ for the *Partition*.

## V. EXPERIMENTS

We now evaluate our approach using the real AOL query logs [14]. We compared our method *Clump* with the state-of-the-art transaction anonymization method *Partition* [8]. The implementation of both algorithms was done in Visual C++ and the experiments were performed on a system with core-2 Duo 2.99GHz CPU with 3.83 GB memory.

### 1) Experiment Setup

**Dataset information** The AOL query log collection dataset consists of 20M web queries collected from 650k users over three months in form of {*AnonID*, *QueryContent*, *QueryTime*, *ItemRank*, *ClickURL*} and are sorted by anonymous *AnonID* (user ID). Our experiments focused on anonymizing *QueryContent*. The dataset has a size of 2.2GB and is divided into 10 subsets, each of which has similar characteristics and size. In our experiment, we used the first subset. In addition, we merged the queries issued by the same *AnonID* into one transaction because each query is too short, and removed duplicate items, resulting in 53,058 queries or transactions with the average transaction length of 20.93.

We generated the item taxonomy $T$ using the *WordNet* dictionary [5]. According to the *WordNet*, each noun has multiple senses. A sense is represented by a synset, i.e., a set of words with the same meaning. We used the first word to represent a synset. In pre-processing the AOL dataset, we discarded words that are not in the *WordNet* dictionary. We treated each noun as an item and interpreted each noun by its most frequently used sense i.e., the first synset. Therefore, nouns together with the *is-a-kind-of* links among them comprise a tree. The generated taxonomy tree contains 25645 items and has the height 18.

We investigate the following four quality indicators: a) distortion (i.e., information loss), b) average generalized transaction length, which reflects the number of items suppressed, c) average level of generalized items (with the root at level 1), and d) execution time. The distortion is measured by $\Sigma GGD(S_i, LCG(S_i))$ over the clusters $S_i$ for *Clump*, and by $\Sigma GGD(S_i, t_j)$ over the sub-partitions $S_i$ for *Partition* where $t_j$ is the generalized transaction.

**Parameters** The first parameter is the anonymity parameter $k$. We set $k$ to 5, 7, 10, and 15. Another parameter is the database size $|D|$ (i.e., the number of transactions). In our experiments, we used the first 1000, 10000, and 53,058 transactions to evaluate the runtime and the effect of "transaction density" on our algorithm performance. The transaction density is measured by the ratio $N_{total} / (|D| \times |L|)$, where $N_{total}$ is the sum of number of items in all transactions, $|D|$ is the number of transactions, and $|L|$ is the number of leaf items in our taxonomy. $|D| \times |L|$ is the maximum possible number of items that can occur in $|D|$ transactions. Table III shows the density of the first $|D|$ transactions. Clearly, a database gets sparser as $|D|$ grows. Unless otherwise stated, we set the parameter $r$=10 (a parameter used by *Clump*).

#### Table III
#### Transaction database density

| $|D|$ | 1,000 | 10,000 | 20,000 | 30,000 | 40,000 | 53,058 |
|---|---|---|---|---|---|---|
| Density | 0.28% | 0.25% | 0.20% | 0.16% | 0.14% | 0.11% |

### 2) Results

As discussed in Section I.A, one of our goals is to preserve duplicate items after generalization because duplication of items tells some information about the number of items in an original transaction, which is useful to data analysis. To study the effectiveness of achieving this goal, we consider two versions of the result produced by *Clump*, denoted by *Clump1* and *Clump2*. *Clump1* represents the result produced by *Clump* as discussed in Section IV, thus, preserves duplicate items in *LCG*. *Clump2* represents the result after removing all duplicate items from *LCG*.

Figures 3, 4, 5 show the results with respect to information loss, average transaction length, and average level of generalized items. Below, we discuss each in details.

**Information loss** Fig. 3 clearly shows that the information loss is reduced by the proposed *Clump* compared with *Partition*. The reduction is as much as 30%. As we shall see shortly, this reduction comes from the lower generalization level of the generalized items in *LCG*, which comes from the effectiveness of grouping similar transactions in our clustering algorithm. However, the difference between *Clump1* and *Clump2* is very small.

A close look reveals that many duplicate items preserved by *Clump1* are at a high level of the taxonomy tree. For such items, generalization has a *GGD* close to that of suppressing an item. However, this does not mean that such duplicate items carry no information. Indeed, duplicates of items tell some information about the quantity or frequency of an item in an original transaction. Such information is not modeled by the *GGD* metric.

As the database gets larger, the data gets sparser; the improvement of *Clump* over *Partition* gets smaller. In fact, when data is too sparse, no algorithm is expected to perform well. As the privacy parameter $k$ increases, the improvement reduces. This is because each cluster contains more transactions, possibly of different lengths; therefore, more generalization and more suppression are required for the *LCG* of such clusters. Typically, $k$ in the range of [5,10] would provide adequate protection.

**Average generalized transaction length** Fig. 4 shows the average length of generalized transactions. *Clump1* has significantly larger length than *Clump2* and *Partition*. This longer transaction length is mainly the consequence of preserving duplicate items in *LCG* by *Clump1*. As discussed above, duplicate items carry useful information about the quantity or frequency of items in an original transaction. The proposed *Clump* preserves better such information than *Partition*.

**Average level of generalized items** Fig. 5 shows that the average level of generalized items for *Clump2* is lower than that for *Partition* which is lower than that for *Clump1* (recall that the root item is at level 1). This is due to the fact that many duplicate items preserved by *Clump1* are at a level close to the root. When such duplicates are removed (i.e., *Clump2*), the remaining items have a lower average level than *Partition*.

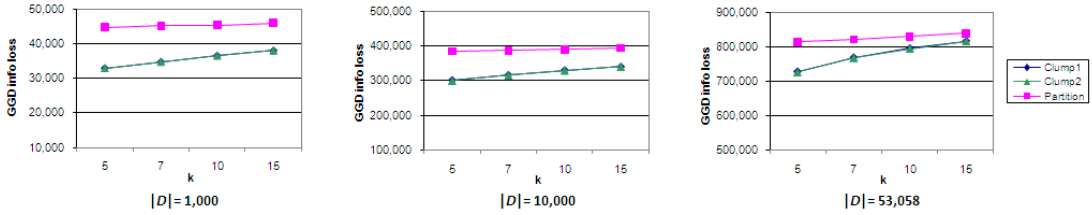

Fig. 3. Comparison of information loss

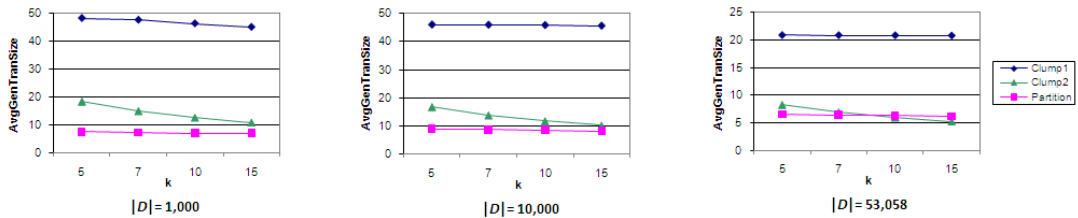

Fig. 4. Comparison of average generalized transaction length

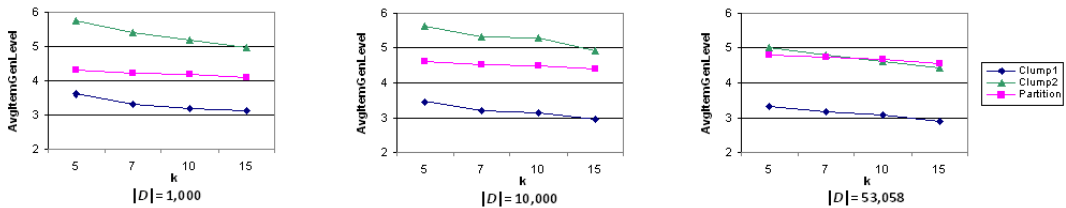

Fig. 5. Comparison of average level of generalized item

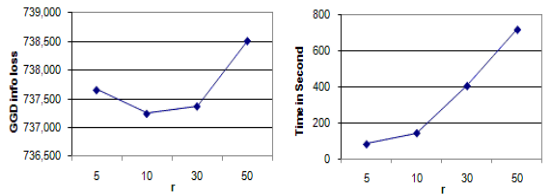

Fig. 6. Effect of *r* on *Clump1*

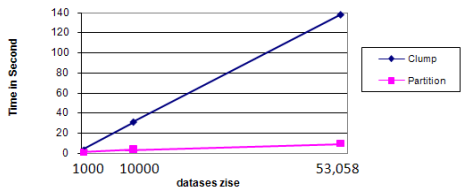

Fig. 7. Comparison of running time

**Sensitivity to the parameter** *r* This is the number of top clusters examined for assigning each transaction. A larger *r* means that more clusters will be examined to assign a transaction, thus, a better local optimal cluster but a longer runtime. In this experiment, we set $|D|$=53,058 and $k$=5. As shown in Fig. 6, we set *r* to 5, 10, 30, 50, and 100. This experiment shows that a larger *r* does not always give a better result since *Clump* works in a greedy manner and by increasing the number of clusters to examine, we may come up with a locally optimal choice that later increases the overall information loss. Our experiments show that setting $r$=10 achieves a good result.

**Runtime** Fig. 7 depicts the runtime comparison for $k$=5 and $r$=10. *Clump* takes longer time than *Partition* does. In fact, the small runtime of *Partition* is largely due to the fact that the top-down algorithm stops partitioning the data at a high level of the taxonomy because a sub-partition contains less than $k$ transactions. Thus, this small runtime is in fact the costly information loss. *Clump* takes a longer runtime but is still linearly scalable with respect to the data size. Considering the notably less information loss, the longer runtime of *Clump* is justified.

## VI. RELATED WORK

A recent survey [4] discussed seven query log privacy-enhancing techniques from a policy perspective, including deleting entire query logs, hashing query log content, deleting user identifiers, scrubbing personal information from query content, hashing user identifiers, shortening sessions, and deleting infrequent queries. Although these techniques protect privacy to some extent, there is a lack of formal privacy guarantees. For example, the release of the AOL query log data still leads to the re-identification of a search engine user even after hashing user's identifiers [2]. The challenge is that the query content itself may be used together with publicly available information for linking attacks.

In *token based hashing* [10] a query log is anonymized by tokenizing each query term and securely hashing each token to an identifier. However, if an unanonymized reference query log has been released previously, the adversary could employ the reference query log to extract statistical properties of query terms in the log and then processes the anonymized log to invert the hash function based on co-occurrences of tokens within queries.

*Secret sharing* [1] is another method which splits a query into $k$ random shares and publishes a new share for each distinct user issuing the same query. This technique guarantees $k$-anonymity because each share is useless on its own and all the $k$ shares are required to decode the secret. This means that a query can be decoded only when there are at least $k$ users issuing that query. The result is equivalent to suppressing all queries issued by less than $k$ users. Since queries are typically sparse, many queries will be suppressed as a result.

*Split personality*, also proposed in [1], splits the logs of each user on the basis of "interests" so that the users become dissimilar to themselves, thus reducing the possibility of reconstructing a full user trace (i.e. search history of a user). This distortion also makes it more difficult for researchers to correlate different facets.

The work on transaction anonymization is studied in the database and data mining communities. Other than the *Partition* algorithm [8] we discussed in Section I.A, some techniques such as $(h; k; p)$-coherence [19], using suppression technique, and $k^m$-anonymity [18], using generalization, have been proposed. Both works assume that a realistic adversary is limited by a maximum number of item occurrences that can be acquired as background knowledge. As pointed out in [8], if background knowledge can be on the absence of items, the adversary may exclude transactions using this knowledge and focus on fewer than $k$ transactions. The $k$-anonymity avoids this problem because all transactions in the same equivalence class are identical.

## VII. CONCLUSION

The objective of publishing query logs for research is constrained by privacy concerns and it is a challenging problem to achieve a good tradeoff between privacy and utility of query log data. In this paper, we proposed a novel solution to this problem by casting it as a special clustering problem and generalizing all transactions in each cluster to their least common generalization (*LCG*). The goal of clustering is to group transactions into clusters so that the overall distortion is minimized and each cluster has at least the size $k$.

We devised efficient algorithms to find a good clustering. Our studies showed that the proposed algorithm retains a better data utility in terms of less data generalization and preserving more items, compared to the state-of-the-art transaction anonymization approaches.

ACKNOWLEDGMENT

Authors would like to thank Junqiang Liu for his assistance in implementation and also reviewers of SECRYPT 2010 conference for their feedback.